\documentclass[pr,cha,onecolumn,superscriptaddress,preprintnumbers,showpacs, amsmath,amssymb]{revtex4-1}
\usepackage{bm}
\usepackage[colorlinks=true,linkcolor=blue,citecolor=blue]{hyperref}
\usepackage{amsmath}
\usepackage{amssymb}
\usepackage{amsthm}
\usepackage{amsfonts}
\usepackage{enumerate}
\usepackage{latexsym}
\usepackage{ifpdf}
\usepackage{graphicx}
\usepackage{makeidx}
\expandafter\ifx\csname package@font\endcsname\relax\else
\expandafter\expandafter
\expandafter\usepackage
\expandafter\expandafter
\expandafter{\csname package@font\endcsname}
\fi
\usepackage{color}
\usepackage{times}
\usepackage{ulem}

\begin{document}

\title{Electronic and magnetic properties of Lu and LuH$_2$}

\author{Shunda Zhang}
\thanks{These authors contributed equally to this work}
\affiliation{Ningbo Institute of Materials Technology and Engineering, Chinese Academy of Sciences, Ningbo 315201, China}
\author{Jiachang Bi}
\thanks{These authors contributed equally to this work}
\affiliation{Ningbo Institute of Materials Technology and Engineering, Chinese Academy of Sciences, Ningbo 315201, China}
\author{Ruyi Zhang}
\thanks{These authors contributed equally to this work}
\affiliation{Ningbo Institute of Materials Technology and Engineering, Chinese Academy of Sciences, Ningbo 315201, China}
\author{Peiyi Li}
\affiliation{Ningbo Institute of Materials Technology and Engineering, Chinese Academy of Sciences, Ningbo 315201, China}
\author{Fugang Qi}
\affiliation{Ningbo Institute of Materials Technology and Engineering, Chinese Academy of Sciences, Ningbo 315201, China}
\author{Zhiyang Wei}
\affiliation{Ningbo Institute of Materials Technology and Engineering, Chinese Academy of Sciences, Ningbo 315201, China}
\author{Yanwei Cao}
\email{ywcao@nimte.ac.cn}
\affiliation{Ningbo Institute of Materials Technology and Engineering, Chinese Academy of Sciences, Ningbo 315201, China}
\affiliation{Center of Materials Science and Optoelectronics Engineering, University of Chinese Academy of Sciences, Beijing 100049, China}

\date{\today}

\begin{abstract}

Clarifying the electronic and magnetic properties of lutetium, lutetium dihydride, and lutetium oxide is very helpful to understand the emergent phenomena in lutetium-based compounds (such as superconductivity-like transitions near room temperature). However, this kind of study is still scarce at present. Here, we report on the electronic and magnetic properties of lutetium metals, lutetium dihydride powders, and lutetium oxide powders. Crystal structures and chemical compositions of these samples were characterized by X-ray diffraction and X-ray photoemission spectroscopy, respectively.  Electrical transport measurements show that the resistance of lutetium has a linear behavior depending on temperature, whereas the resistance of lutetium dihydride powders changes little with decreasing the temperature. More interestingly, paramagnetism-ferromagnetism-spin glass transitions were observed at near 240 and 200 K, respectively, in lutetium metals. To understand their origins, the measurement of inductively coupled plasma optical emission spectroscopy was performed, revealing that the concentrations of dilute magnetic impurities in lutetium and lutetium dihydride are around 0.01\%, which can lead to the presence of spin glassy behavior. Our work uncovers the complex magnetic properties of lutetium and lutetium dihydride and suggests their close connections to the electronic and magnetic transitions of Lutetium-based compounds (such as Lu-H-N).

\end{abstract}

\maketitle
\newpage

\section{Introduction}

Due to the discovery of room-temperature superconductivity in nitrogen-doped lutetium hydrides (Lu-H-N), the study of Lu-based compounds attracts a great deal of attention very recently \cite{Nature-2023-Dias,SCPMA-2023-Jin,arXiv-2023-LiuP, CPL-2023-Cheng, arXiv-2023-XM,CPL-2023-LiuM,arXiv-2023-Cui,arXiv-2023-Ho,arXiv-2023-Zurek,arXiv-2023-YZ,arXiv-2023-LiuX,arXiv-2023-Sun,arXiv-2023-Cheng,SB-2023-Cheng,arXiv-2023-MM,arXiv-2023-Boeri,arXiv-2023-Heil,arXiv-2023-LuT,arXiv-2023-BM,arXiv-2023-PL,arXiv-2023-XT,JSNM-2023-VA,arXiv-2023-ZL,arXiv-2023-DD,ArXiv-2020-Duan,R-2022-Cui,I-2021-Cui,M-2021-W,arXiv-2023-Hirsch,arXiv-2023-Harshman}. Theoretical predication of high-temperature superconductivity in Lu hydrides ($\sim$ 273 K at the pressure 100 GPa in LuH$_6$) has been reported in the year 2020 \cite{ArXiv-2020-Duan,R-2022-Cui}. Experimental observation of superconductivity (with the critical temperature T$_C$ $\sim$ 15 K at the pressure 128 GPa) was realized in $ Fm\bar{3}m $-LuH$_3$ in 2021 \cite{I-2021-Cui,M-2021-W}. Then, the superconductivity critical temperature was increased to 71 K at the pressure 218 GPa in the compound Lu$_4$H$_{23}$ \cite{SCPMA-2023-Jin}. In the Lu-H-N system with room-temperature superconductivity-like transitions, one of the most intriguing phenomena is the pressure-induced color from dark-blue to red \cite{Nature-2023-Dias}, which has been observed by different groups and explained to be the shift of the minimum of optical reflectivity under pressure \cite{CPL-2023-Cheng,arXiv-2023-YZ,arXiv-2023-LiuX,SB-2023-Cheng,arXiv-2023-BM,arXiv-2023-XT,arXiv-2023-ZL,arXiv-2023-DD}. However, whether near-ambient superconductivity can exist in Lu-based compounds is still an open question \cite{Nature-2023-Dias,SCPMA-2023-Jin,arXiv-2023-LiuP, CPL-2023-Cheng, arXiv-2023-XM,CPL-2023-LiuM,arXiv-2023-Cui,arXiv-2023-Ho,arXiv-2023-Zurek,arXiv-2023-YZ,arXiv-2023-LiuX,arXiv-2023-Sun,arXiv-2023-Cheng,SB-2023-Cheng,arXiv-2023-MM,arXiv-2023-Boeri,arXiv-2023-Heil,arXiv-2023-LuT,arXiv-2023-BM,arXiv-2023-PL,arXiv-2023-XT,JSNM-2023-VA,arXiv-2023-ZL,arXiv-2023-DD,ArXiv-2020-Duan,R-2022-Cui,I-2021-Cui,M-2021-W}. 

\begin{figure*}[b]
\includegraphics[width=0.85\textwidth]{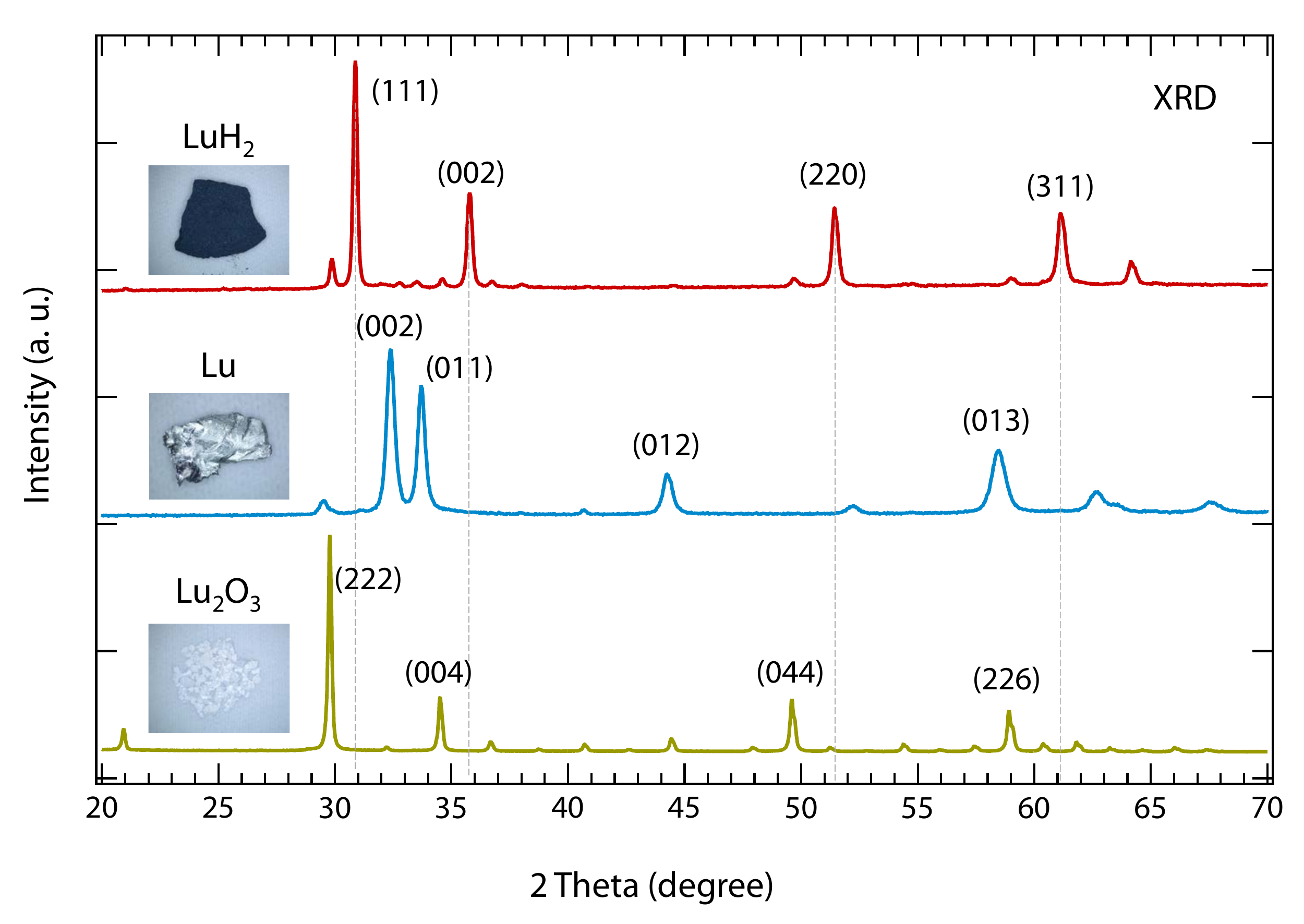}
\caption{\label{}  XRD data of Lu metals, LuH$_2$ powders, and Lu$_2$O$_3$ powders at room temperature. The insets show photographs of three samples. The colours of Lu metals, LuH$_2$ powders (compressed), and Lu$_2$O$_3$ powders are slivery white, dark blue, and white, respectively.}
\end{figure*}

Besides the superconductivity, several other intriguing properties were also reported in Lu-based compounds such as hydrogen storage in LuH$_2$ \cite{IJHE-2021-Ouadah,Vacuum-2023-Saidi}, high electrical conductivity, and strong spin-orbit coupling in lutetium monoxide LuO \cite{ACSOmega-2018-Fukumura}, phase transformations in the Lu and LuH$_3$ \cite{PRB-1998-Vohra,JAC-2007-Palasyuk}, ultrawide bandgap (5.5-5.9 eV) and luminescence in Lu$_2$O$_3$ \cite{APL-2021-Zhang,OME-2020-AKIHIKO,arXiv-2023-MM}. The element Lu with electron configuration 4f$^{14}$5d$^1$6s$^2$ is the last one of 14 rare-earth elements. Generally, the valence states of Lu can be Lu$^{0}$, Lu$^{2+}$, and Lu$^{3+}$ at ambient. Interestingly, superconductivity was observed in the compounds with Lu$^{0}$ or Lu$^{2+}$ or Lu$^{3+}$ valence state \cite{I-2021-Cui,Nature-2023-Dias}. Clarifying the electronic and magnetic properties of lutetium-based compounds is very helpful to understand the emergent phenomena. However, this kind of study is still scarce at present \cite{PR-1967-Lee,JCP-1973-Sped}.

To address the above, we investigated the structural, electronic, and magnetic properties of Lu$^{0}$ metals, (Lu$^{2+}$)H$_{2}$ powders, and (Lu$^{3+}$)$_2$O$_{3}$ powders with X-ray diffractometer (XRD), X-ray photoemission spectroscopy (XPS), electrical transport, and a magnetometer with the superconducting quantum interference device (SQUID). It is revealed that both Lu and LuH$_2$ show complex magnetic transitions, which can result from the existence of dilute magnetic impurities, as shown by the measurements of inductively coupled plasma optical emission spectroscopy (ICP-OES).

\begin{figure*}[]
\includegraphics[width=0.95\textwidth]{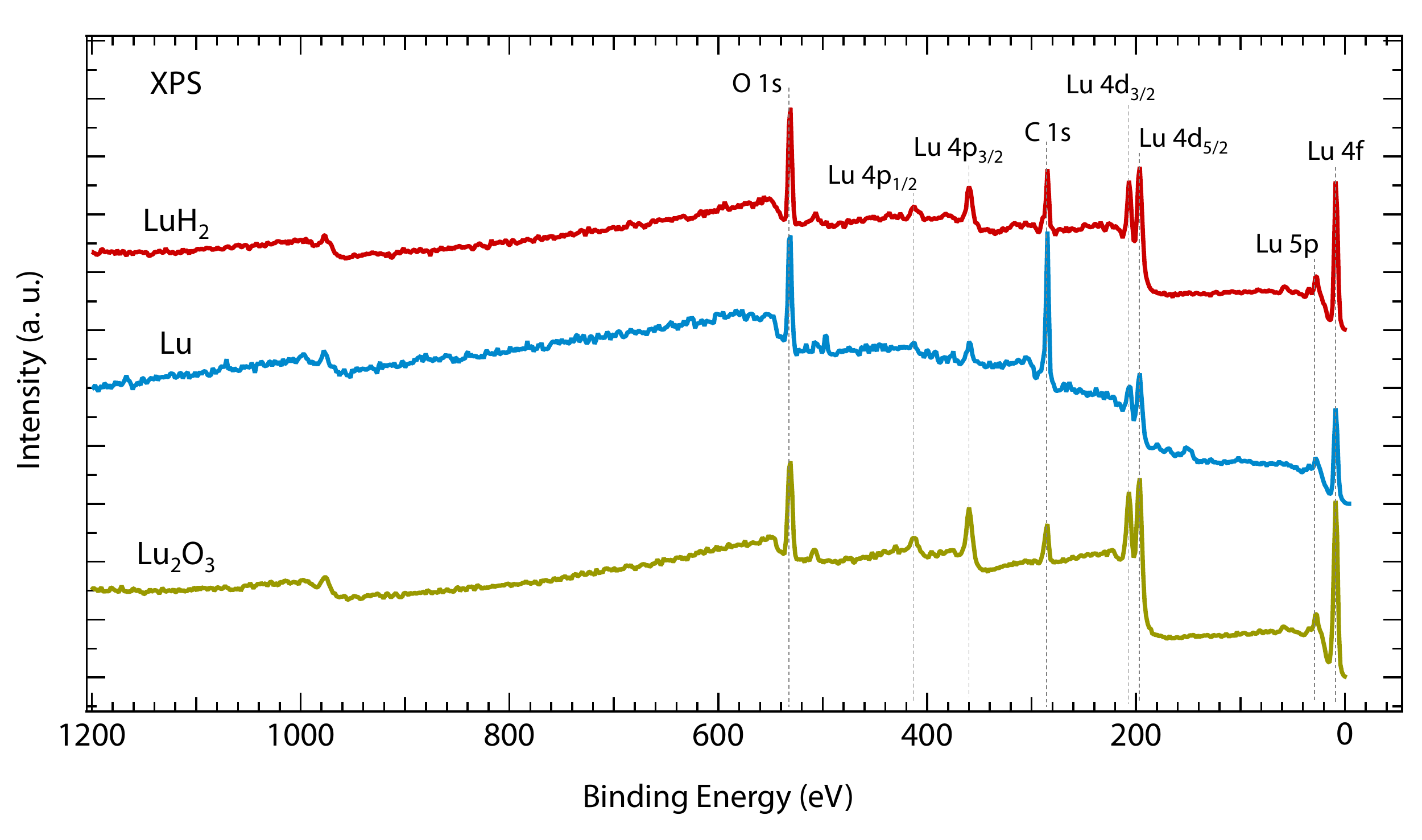}
\caption{\label{} XPS spectra of Lu metals, LuH$_2$ powders, and Lu$_2$O$_3$ powders from 0 to 1200 eV at room temperature.}
\end{figure*}

\section{Experiments}

All samples (Lu polycrystalline metals, LuH$_2$ powders, and Lu$_2$O$_3$ powders, see the insets in Fig.1) studied in this work are commercial. The Lu polycrystalline metals (silvery white, purity of 99.99\%) were ordered from Grirem Advanced Materials (China), whereas the dark blue LuH$_2$ powders (mislabeled as ``Lu'', purity of 99.9\%, analogous to the samples reported in ref. 4) and white Lu$_2$O$_3$ powders were purchased from Macklin (China). The 2$\theta-\omega$ scans on three samples were carried out by a powder XRD (Bruker D8 Discover) with the Cu K$_{\alpha}$ source at room temperature. To characterize the chemical compositions of samples, XPS (monochromatic Al K$_{\alpha}$ radiation, h$\upsilon$ = 1486.6 eV, Kratos AXIS Supra) was performed at room temperature. The electrical properties were measured from 300 to 2 K by Physical Property Measurement System (PPMS) in a van der Pauw geometry (DynaCool, Quantum Design). To measure the electrical transport of  LuH$_2$, it is noted that the LuH$_2$ powders were compressed into several tablets (10 mm in diameter) by a hydraulic press (YLJ-24TA, MTI-KJ group, China) under the pressure 20 MPa at room temperature. Temperature-dependent magnetic properties were characterized by a SQUID magnetometer (Quantum Design) from the temperature 300 to 2 K. To detect magnetic impurities, inductively coupled plasma optical emission spectroscopy (ICP-OES, SPECTRO ARCOS \uppercase\expandafter{\romannumeral2}) was applied. For preparing the ICP-OES measurement, the samples (Lu and LuH$_2$) were digested in hot 36 $\%$ hydrochloric acid and 65 $\%$ nitric acid, and the solutions were subsequently diluted to appropriate concentrations after complete dissolution of samples.

\section{Results and Discussion}

First, we characterized the crystal structures of these three samples. As shown in Fig. 1, the diffraction peaks of these three samples are totally different. The dominative feature of LuH$_2$ XRD data (see the red curve in Fig. 1) is the four main diffraction peaks in the range 20-70 degrees, which can be assigned to (111), (002), (220), and (311) diffraction peaks, respectively \cite{arXiv-2023-Cheng}. The crystal structure of LuH$_2$ is cubic (Fm$\bar{3}$m space group) with an estimated lattice parameter $a$ $\sim$ 5.02\AA ~ \cite{Vacuum-2023-Saidi}, whereas the lattice parameters of hexagonal Lu (P6$_3$/mmc) are $a$ = $b$ $\sim$ 3.516 \AA ~ and $c$ $\sim$  5.573 \AA ~ \cite{PRB-1998-Vohra}. The four main peaks of Lu XRD data (see blue curve in Fig. 1) are (002), (011), (012), and (013) diffraction peaks of polycrystalline Lu. For the compound Lu$_2$O$_3$, its crystal structure is cubic with the lattice parameter $a$ $\sim$ 10.36 \AA, also showing four main diffraction peaks (222), (004), (044), and (226) \cite{Growth-2014-Guzik}. It is noteworthy that several small peaks can be seen in the XRD curve of LuH$_2$  (see red curve in Fig. 1). Compared to the XRD lineshapes of Lu and Lu$_2$O$_3$ (see blue and brown curves in Fig. 1), it is indicated that the LuH$_2$ sample has little Lu and Lu$_2$O$_3$ compositions, which is consistent with the previous report \cite{CPL-2023-Cheng}.

\begin{figure*}[]
\includegraphics[width=0.95\textwidth]{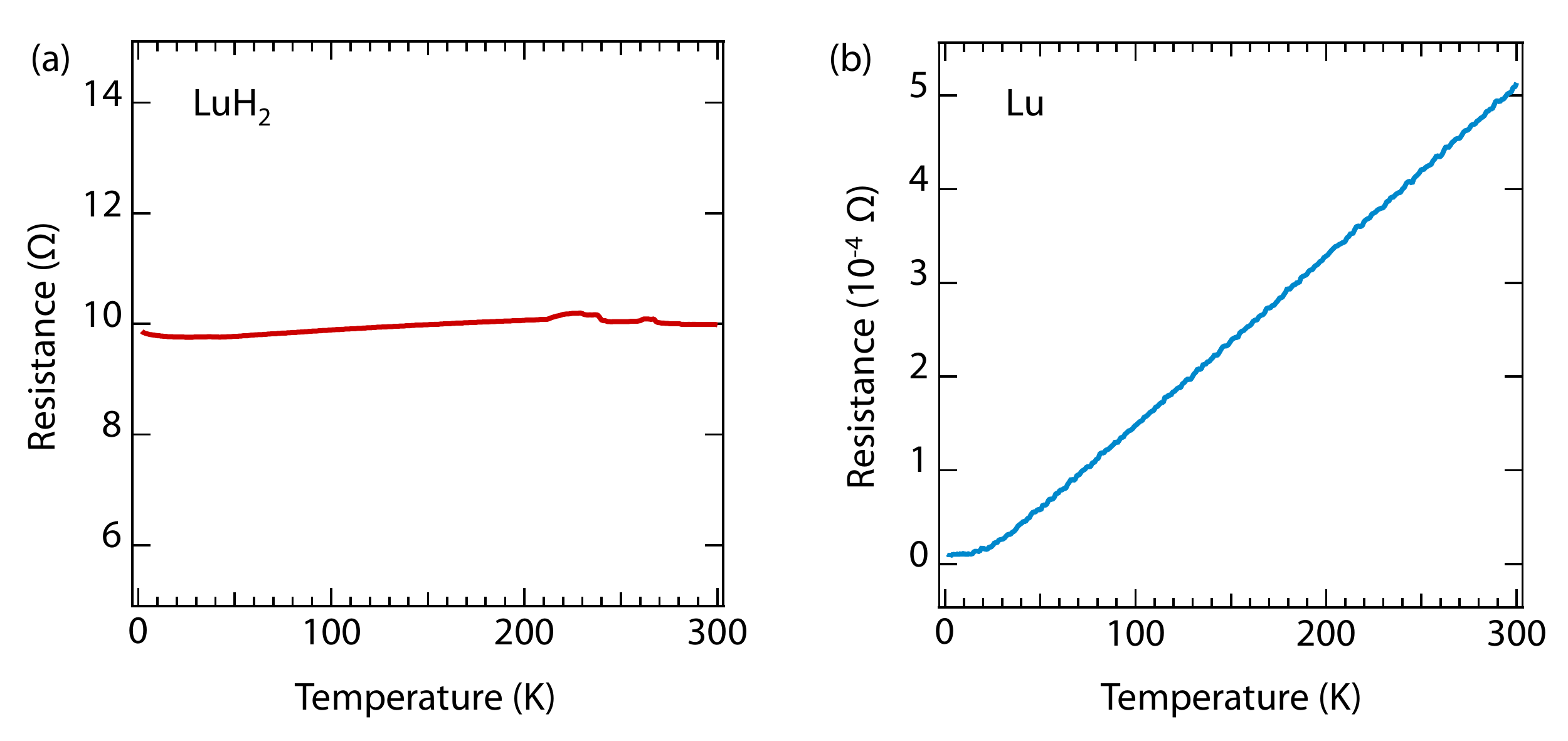}
\caption{\label{} Temperature-dependent resistances of (a) LuH$_2$ powders (compressed) and (b) Lu metals from 300 to 2 K.}
\end{figure*}

\begin{figure*}[th]
	\includegraphics[width=0.95\textwidth]{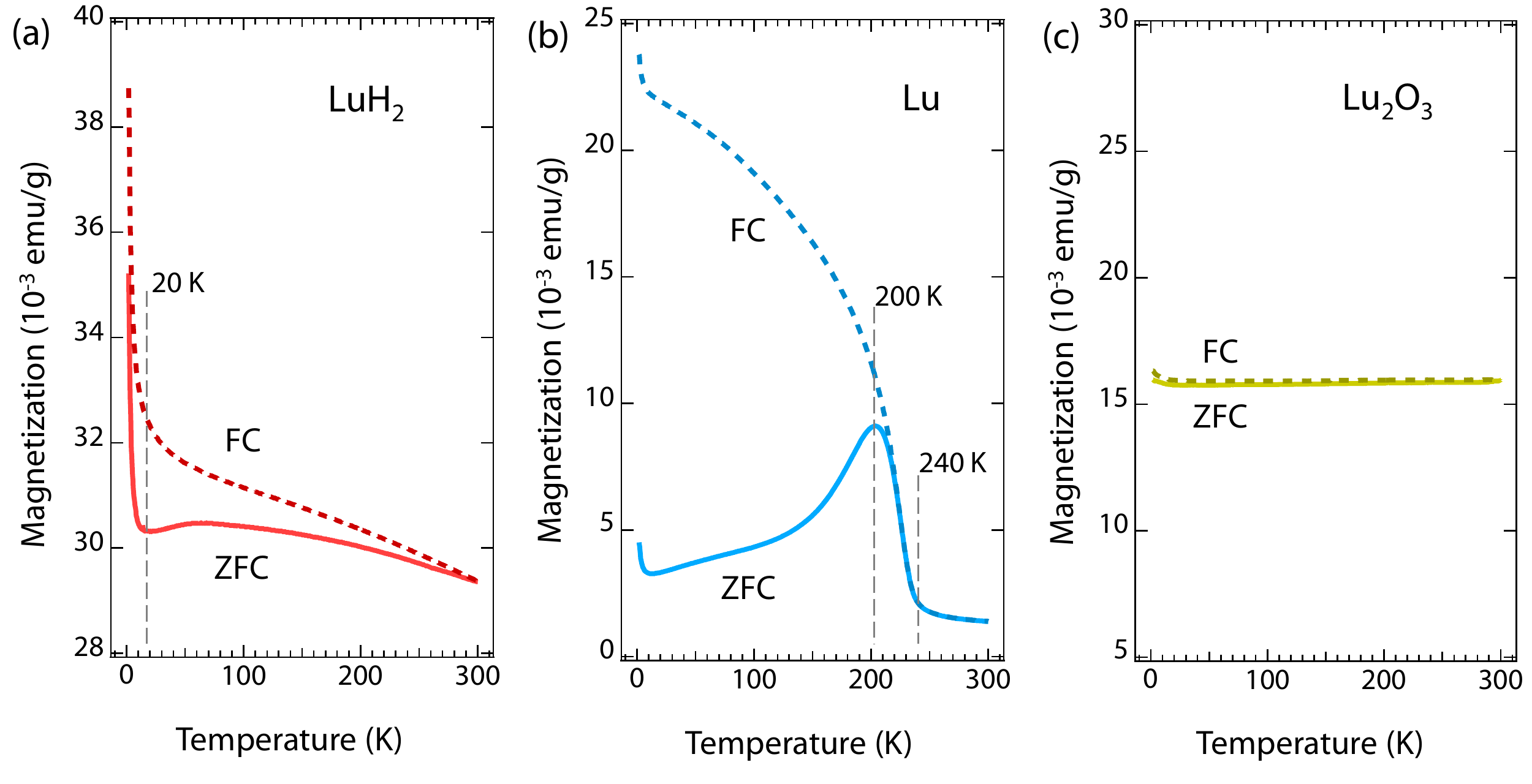}
	\caption{\label{} Temperature-dependent magnetizations (ZFC and FC) of (a) LuH$_2$ powders  (compressed), (b) Lu metals, and (c) Lu$_2$O$_3$ powders from 300 to 2 K. During measurements, the cooling field was set at 1000 Oe.}
\end{figure*}

To further investigate the chemical compositions of Lu, LuH$_2$, and Lu$_2$O$_3$ samples, we performed XPS at room temperature, which is a surface-sensitive technique to probe the valence states of compounds. As seen in Fig. 2, apart from the oxygen (O 1s $\sim$ 531 eV) and carbon (C 1s $\sim$ 284.8 eV)  existed on sample surfaces, no distinct impurity is detected, verifying the purity of these samples. It is noted that the detection limits in XPS is 1\%-0.1\% generally \cite{SIA-2014-AS}. Interestingly, a number of peaks can be observed in the XPS spectra, most of which are assigned to Lu core-level states such as Lu 4$f$ ($\sim$ 9 eV), 5$p$ ($\sim$ 28 eV), 4$d$$_{5/2}$ ($\sim$ 196 eV), 4$d$$_{3/2}$ ($\sim$ 207 eV), 4$p$$_{3/2}$ ($\sim$ 360 eV), and 4$p$$_{1/2}$ ($\sim$ 411 eV) states \cite{JETP-2013-Kai}. Here, we emphasize that the surfaces of both Lu and LuH$_2$ are easily oxidized.

Next, we measured the electronic properties of Lu and LuH$_{2}$ samples by electrical transport from 300 to 2 K. As shown in Fig. 3, both Lu (size $\sim$ 3.7 $\times$ 1.9 $\times$ 0.5 mm$^{3}$) and LuH$_{2}$ (size $\sim$ 5.1 $\times$ 6.1 $\times$ 0.8 mm$^{3}$) samples are highly conducting, the resistances of which are $\sim$ 10 $\Omega$ and $\sim$ 10$^{-4}$ $\Omega$, respectively. With decreasing the temperature from 300 to 2 K, the resistance of LuH$_2$ changes little, whereas the resistance of Lu shows a strong linear behavior down to 20 K, then becomes a constant below 20 K. The resistance of Lu changed from 5.13 $\times$ 10$^{-4}$ $\Omega$ at 300 K to 0.1 $\times$ 10$^{-4}$ $\Omega$ at 20 K. It is noted that the data in  Fig. 3 (a) is noisy due to the measured LuH$_2$ sample in Fig. 3 (a) being a compressed tablet from LuH$_2$ powders.

At last, to investigate the magnetic properties of Lu, LuH$_{2}$, and Lu$_2$O$_3$, we measured the zero-field cooling (ZFC) and field cooling (FC) magnetization curves by SQUID. During measurements, the cooling field was set at 1000 Oe. As seen in Fig. 4, Lu,  LuH$_{2}$, and Lu$_2$O$_3$ present different magnetic properties. For LuH$_{2}$, the magnetization under FC protocol (see the dashed red curve in Fig. 4 (a)) increases monotonously with decreasing the temperature to 20 K, whereas the magnetization has a broad bump for the ZFC protocol (see the solid red curve in Fig. 4 (a)). Comparing to LuH$_{2}$ in Fig. 4 (a), the behavior of temperature-dependent magnetization of Lu is analogous but more clear. As shown in Fig. 4 (b), there is a distinct separation between ZFC and FC curves, which can result from the spin glass transition \cite{Book-1993-AM,JPCM-2008-Huang,PRB-2009-Wang}. With decreasing the temperature from 300 to 2 K, two critical features ( at $\sim$ 240 K and $\sim$ 200 K) can be observed. The first one $\sim$ 240 K can be interpreted as the paramagnetic to the ferromagnetic phase transition, whereas the second one $\sim$ 200 K is due to the spin glass transition.  In contrast to Lu and LuH$_{2}$, the magnetizations of Lu$_2$O$_3$ are almost independent of temperature, indicating the diamagnetic nature of Lu$_2$O$_3$ \cite{MCP-2010-Ben}.

\begin{figure*}[]
	\includegraphics[width=0.8\textwidth]{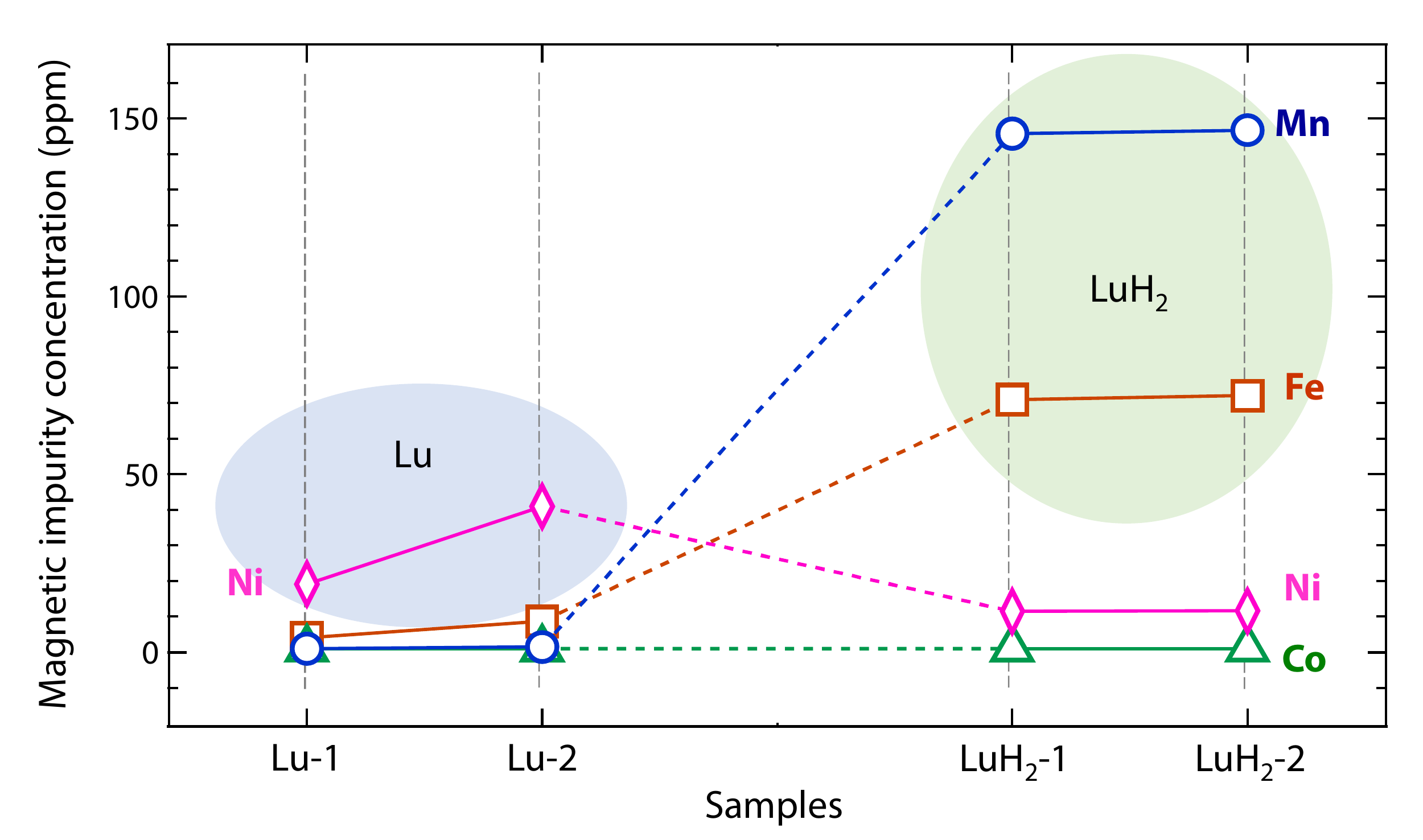}
	\caption{\label{} Magnetic impurity concentration (mass) of Mn, Fe, Co, Ni elements in Lu and LuH$_2$ compounds, which was determined by ICP-OES. Here, Lu-1 and Lu-2 are the labels of two Lu samples, whereas LuH$_2$-1 and LuH$_2$-2 are the marks of two LuH$_2$ specimens.}
\end{figure*}

Then, we discuss the origins of complex magnetic transitions. Generally, there are two scenarios that can explain the paramagnetic-ferromagnetic-spin glass transitions in Lu and LuH$_2$ compounds, which are self-induced spin glass states and dilute magnetic impurities-induced spin glass states, respectively \cite{Book-1993-AM,Science-2020-Kam}. For the first scenario, it is noted that a self-induced spin glass state has been observed in crystalline neodymium very recently, resulting from valley-like pockets of degenerate magnetic wave vectors \cite{Science-2020-Kam}. For the second one, the spin glassy behavior in dilute magnetic alloys has been studied for a long time \cite{JPFMP-1975-PW}. It has been reported that dilute Fe impurity even at the level of 10 atomic parts per million (ppm) can lead to the formation of spin glassy behavior \cite{PRL-1984-Peter}. Therefore, to detect the concentration of magnetic impurities (such as Mn, Fe, Co, Ni) in Lu and LuH$_2$ compounds at the level of ppm, ICP-OES was performed at room temperature. As shown in Fig. 5, the purity of Lu with the magnetic impurity concentration near 30 ppm is much higher than LuH$_2$ with the concentration near 150 ppm. In addition, the main impurity of Lu is Ni element, whereas they are Mn and Fe elements in LuH$_2$. The existence of non-negligible magnetic impurities in Lu and LuH$_2$ can lead to the emergence of complex magnetic transitions and spin glassy behavior.

More interestingly, the critical temperatures (200-240 K) of magnetic transitions observed in this work are pretty close to the electronic and magnetic transitions reported in LuH$_2$ and Lu-H-N compounds \cite{CPL-2023-Cheng,arXiv-2023-XM,arXiv-2023-LiuX,Nature-2023-Dias}. For example, temperature-dependent resistance of LuH$_2$ at the pressure 0.5 (0.7) GPa shows a distinct peak near 230 K \cite{CPL-2023-Cheng}. Analogous features at temperatures 200-250 K can also be observed in the Lu-H-N system by different groups \cite{arXiv-2023-XM,arXiv-2023-LiuX}. Furthermore, the data of the magnetic susceptibility of the Lu-H-N system also show the presence of magnetic transitions at temperatures 200-250 K \cite{Nature-2023-Dias}. These above phenomena strongly suggest the close connection of the spin glass transition to the electronic and magnetic transitions of the Lu-H-N system \cite{JSNM-2023-VA}.

\section{Conclusion}

In this work, we characterized the electronic and magnetic properties of Lu, LuH$_2$, and Lu$_2$O$_3$ by XRD, XPS, electrical transport, and SQUID. Both Lu and LuH$_2$ are metallic, whereas Lu$_2$O$_3$ is insulating. The resistance of Lu has a linear behavior depending on temperature, whereas the resistance of LuH$_2$ changes little. More interestingly, the magnetizations of both Lu and LuH$_2$ show a spin glass feature, which can result from the existence of dilute magnetic impurities with near 0.01\% concentration. Our work uncovers the complex magnetic properties of Lu-based compounds and suggests their close connections to the electronic and magnetic transitions of the Lu-H-N system.

\section{Acknowledgments}

We thank Xinming Wang, Jie Sun, Kemin Jiang, Yusha Du, Kai Shen, Li Wang and Chuyi Ning for helping experimental setup. We acknowledge insightful discussions with Prof. Jiandong Guo, Prof. Aihua Sun, Prof. Baomin Wang, Dr. Shaozhu Xiao, Prof. Aru Yan, Prof. Rende Chen, and Prof. Israel Felner. This work was supported by the National Key R\&D Program of China (Grant No. 2022YFA1403000), the National Natural Science Foundation of China (Grant Nos. U2032126 and 11874058), the Pioneer Hundred Talents Program of the Chinese Academy of Sciences, the Zhejiang Provincial Natural Science Foundation of China under Grant No. LXR22E020001, the Beijing National Laboratory for Condensed Matter Physics, the Ningbo Natural Science Foundation (Grant No. 20221JCGY010338), and the Ningbo Science and Technology Bureau (Grant No. 2022Z086).

\section{Author declarations}

\subsection{Conflict of Interest}
The authors have no conflicts to disclose.

\subsection{Author Contributions}
Shunda Zhang: Data curation (equal); Formal analysis (equal); Investigation (equal); Validation (equal); Writing – original draft (equal). Jiachang Bi: Data curation (equal); Formal analysis (equal); Investigation (equal); Validation (equal); Writing – original draft (equal). Ruyi Zhang: Data curation (equal); Formal analysis (equal); Investigation (equal); Validation (equal); Writing – original draft (equal); Peiyi Li: Formal analysis (equal); Fugang Qi: Formal analysis (equal); Zhiyang Wei: Formal analysis (equal). Yanwei Cao: Conceptualization (lead); Funding acquisition (lead); Validation (equal); Supervision (lead); Writing – review \& editing (equal).

\section{Data Availability}

The data that support the findings of this study are available from the corresponding author upon reasonable request.

\newpage


\begin{thebibliography}{999}

\bibitem{Nature-2023-Dias}
N. Dasenbrock-Gammon, E. Snider, R. McBride, H. Pasan, D. Durkee, N. Khalvashi-Sutter, S. Munasinghe, S. E. Dissanayake, K. V. Lawler, A. Salamat, and R. P. Dias, Evidence of near-ambient superconductivity in a N-doped lutetium hydride, Nature \textbf{615}, 244 (2023).

\bibitem{SCPMA-2023-Jin}
Z. Li, X. He, C. Zhang, K. Lu, B. Min, J. Zhang, S. Zhang, J. Zhao, L. Shi, Y. Peng, S. Feng, Z. Deng, J. Song, Q. Liu, X. Wang, R. Yu, L. Wang, Y. Li, J. D. Bass, V. Prakapenka, S. Chariton, H. Liu, and C. Jin, Superconductivity above 70 K experimentally discovered in lutetium polyhydride, Sci. China Phys. Mech. Astron. \textbf{66}, 267411 (2023).

\bibitem{arXiv-2023-LiuP}
M. Liu, X. Liu, J. Li, J. Liu, Y. Sun, X.-Q. Chen, and P. Liu, On parent structures of near-ambient nitrogen-doped lutetium hydride superconductor, arXiv:2303.06554 (2023).

\bibitem{CPL-2023-Cheng}
P. Shan, N. Wang, X. Zheng, Q. Qiu, Y. Peng, and J. Cheng, Pressure-induced color change in the lutetium dihydride LuH$_2$, Chinese Phys. Lett. \textbf{40}, 046101 (2023).

\bibitem{arXiv-2023-XM}
X. Ming, Y.-J. Zhang, X. Zhu, Q. Li, C. He, Y. Liu, B. Zheng, H. Yang, and H.-H. Wen, Absence of near-ambient superconductivity in LuH$ _ {2\pm\text {x}}$N$ _y$, arXiv:2303.08759 (2023).

\bibitem{CPL-2023-LiuM}
F. Xie, T. Lu, Z. Yu, Y. Wang, Z. Wang, S. Meng, and M. Liu, Lu-H-N phase diagram from first-principles calculations, Chinese Phys. Lett. \textbf{40}, 057401 (2023).

\bibitem{arXiv-2023-Cui}
Z. Huo, D. Duan, T. Ma, Q. Jiang, Z. Zhang, D. An, F. Tian, T. Cui,  First-Principles Study on the Superconductivity of N doped fcc-LuH$_3$, arXiv:2303.12575 (2023).

\bibitem{arXiv-2023-Ho}
Y. Sun, F. Zhang, S. Wu, V. Antropov, and K.-M. Ho, Effect of nitrogen doping and pressure on the stability of cubic LuH$_3$, arXiv:2303.14034 (2023).

\bibitem{arXiv-2023-Zurek}
K. P. Hilleke, X. Wang, D. Luo, N. Geng, B. Wang, and E. Zurek, Structure, Stability and Superconductivity of N-doped Lutetium Hydrides at kbar Pressures, arXiv:2303.15622 (2023).

\bibitem{arXiv-2023-YZ}
Y.-J. Zhang, X. Ming, Q. Li, X. Zhu, B. Zheng, Y. Liu, C. He, H. Yang, H.-H. Wen, Pressure Induced Color Change and Evolution of Metallic Behavior in Nitrogen-Doped Lutetium Hydride, Sci. China Phys. Mech. \textbf{66}, 287411 (2023).

\bibitem{arXiv-2023-LiuX}
X. Xing, C. Wang, L. Yu, J. Xu, C. Zhang, M. Zhang, S. Huang, X. Zhang, B. Yang, X. Chen, Y. Zhang, J.-g. Guo, Z. Shi, Y. Ma, C. Chen, X. Liu,   Observation of Non-Superconducting Phase Changes in LuH$_{2\pm\text{x}}$N$_y$, arXiv:2303.17587 (2023).


\bibitem{arXiv-2023-Sun}
S. Cai, J. Guo, H. Shu, L. Yang, P. Wang, Y. Zhou, J. Zhao, J. Han, Q. Wu, W. Yang, T. Xiang, H.-k. Mao, L. Sun, No Evidence of Superconductivity in the Compressed Sample Prepared from the Lutetium Foil and H$_2$/N$_2$ Gas Mixture, arXiv:2304.03100 (2023).


\bibitem{arXiv-2023-Cheng}
N. Wang, J. Hou, Z. Liu, P. Shan, C. Chai, S. Jin, X. Wang, Y. Long,Y. Liu, H. Zhang, X. Dong, J. Cheng, Percolation-Induced Resistivity Drop in Cold-Pressed LuH$_2$, arXiv:2304.00558 (2023).


\bibitem{SB-2023-Cheng}
X. Zhao, P. Shan, N. Wang, Y. Li, Y. Xu, J. Cheng, Pressure Tuning of Optical Reflectivity in LuH$_2$, Sci. Bull. 68, 883 (2023).

\bibitem{arXiv-2023-MM}
O. Moulding, S. Gallego-Parra, P. Toulemonde, G. Garbarino, P. Derango, P. Giroux, and M.-A. M$\acute{e}$asson, Trigonal to cubic structural transition in possibly N-doped LuH$_3$ measured by Raman and X-ray diffraction, arXiv:2304.04310 (2023).

\bibitem{arXiv-2023-Boeri}
P. P. Ferreira, L. J. Conway, A. Cucciari, S. Di Cataldo, F. Giannessi, E. Kogler, L. T. F. Eleno, C. J. Pickard, C. Heil, L. Boeri, Search for Ambient Superconductivity in the Lu-N-H System, arXiv:2304.04447 (2023).

\bibitem{arXiv-2023-Heil}
R. Lucrezi, P. P. Ferreira, M. Aichhorn, and C. Heil, Temperature and quantum anharmonic lattice effects in lutetium trihydride: stability and superconductivity, arXiv:2304.06685 (2023).

\bibitem{arXiv-2023-LuT}
T. Lu, S. Meng, and M. Liu, Electron-phonon interactions in LuH$_2$, LuH$_3$, and LuN, arXiv:2304.06726 (2023).

\bibitem{arXiv-2023-BM}
S.-W. Kim, L. J. Conway, C. J. Pickard, G. L. Pascut, and B. Monserrat, Microscopic theory of colour in lutetium hydride, arXiv:2304.07326 (2023).

\bibitem{arXiv-2023-PL}
P. Li, J. Bi, S. Zhang, R. Cai, G. Su, F. Qi, R. Zhang, Z. Wei, and Y. Cao, Transformation of hexagonal Lu to cubic LuH$_{2+x}$ single-crystalline films, arXiv:2304.07966 (2023).

\bibitem{arXiv-2023-XT}
X. Tao, A. Yang, S. Yang, Y. Quan, and P. Zhang, Leading components and pressure-induced color changes in lutetium-nitrogen-hydrogen system, arXiv:2304.08992 (2023).

\bibitem{JSNM-2023-VA}
V. P. S. Awana, I. Felner, S. Ovchinnikov, J. W. A. Robinson, Short note on the observation of ambient condition room temperature superconductivity in nitrogen‑doped lutetium hydride, J. Supercond. Nov. Magn. (2023).

\bibitem{arXiv-2023-ZL}
Z. Liu, Y. Zhang, S. Huang, X. Ming, Q. Li, C. Pan, Y. Dai, X. Zhou, X. Zhu, H. Yan, and H.-H Wen, Pressure-induced color change arising from transformation between intra- and inter-band transitions in LuH$_{2\pm{x}}$N$_y$, arXiv:2305.06103 (2023).

\bibitem{arXiv-2023-DD}
D. Dangi\'{c}, P. Garcia-Goiricelaya, Y.-W Fang, J. Iba\~{n}ez-Azpiroz, and I. Errea, Ab initio study of the structural, vibrational and optical properties of potential parent structures of nitrogen-doped lutetium hydride, arXiv:2305.06751 (2023).

\bibitem{ArXiv-2020-Duan}
H. Song, Z. Zhang, T. Cui, C. J. Pickard, V. Z. Kresin, and D. Duan, High T$_c$ superconductivity in Heavy Rare Earth hydrides, Chinses Phys.Lett. \textbf{38}, 107401 (2021).

\bibitem{R-2022-Cui}
M. Du, H. Song, Z. Zhang, D. Duan, and T. Cui, Room-Temperature Superconductivity in Yb/Lu Substituted Clathrate Hexahydrides under Moderate Pressure, Research \textbf{2022}, 9784309 (2022).


\bibitem{I-2021-Cui}
M. Shao, S. Chen, W. Chen, K. Zhang, X. Huang, and T. Cui, Superconducting ScH$_3$ and LuH$_3$ at Megabar Pressures, Inorg. Chem. \textbf{60}, 15330 (2021).

\bibitem{M-2021-W}
D. Wang, Y. Ding, and H.-K. Mao, Future Study of Dense Superconducting Hydrides at High Pressure, Materials \textbf{14}, 7563 (2021).

\bibitem{arXiv-2023-Hirsch}
J. E. Hirsch, Enormous variation in homogeneity and other anomalous features of room temperature superconductor samples: a Comment on Nature 615, 244 (2023), arXiv:2304.00190 (2023).

\bibitem{arXiv-2023-Harshman}
D. R. Harshman and A. T. Fiory, Determining the upper critical magnetic field for N-doped lutetium hydride directly from the source data files in Dasenbrock-Gammon et al., Nature 615, 244 (2023), arXiv:2305.12065 (2023).


\bibitem{IJHE-2021-Ouadah}
O. Ouadah, F. Saidi, M. E. A. Miloudi, O. Ziani, A. Mahmoudi, and S. Nasr, DFT investigation analyzed with data mining technique of rare-earth dihydrides REH$_2$ for hydrogen storage, Int. J. Hydrog. Energy \textbf{46}, 32962 (2021).

\bibitem{Vacuum-2023-Saidi}
F. Saidi, S. Mokhdar, M. Dergal, A. Mahmoudi, A. Kallekh, and H. H. Abd El-Gawad, Ab initio study of the structural, electronic, magnetic, mechanical, optical, and dynamical properties of the rare-earth dihydrides MH$_2$ (M= Yb, Sc, Eu, Y, Lu and Gd), Vacuum \textbf{212}, 112011 (2023).


\bibitem{ACSOmega-2018-Fukumura}
K. Kaminaga, D. Oka, T. Hasegawa, and T. Fukumura, New Lutetium Oxide: Electrically Conducting Rock-Salt LuO Epitaxial Thin Film, ACS omega \textbf{3}, 12501 (2018).


\bibitem{PRB-1998-Vohra}
G. N. Chesnut and Y. K. Vohra, Phase transformation in lutetium metal at 88 GPa, Phys.Rev. B \textbf{57}, 10221 (1998).

\bibitem{JAC-2007-Palasyuk}
M. Tkacz and T. Palasyuk, Pressure induced phase transformation of REH$_3$, Journal of Alloys and Compounds, J. Alloys Compd. \textbf{446}, 593 (2007).

\bibitem{APL-2021-Zhang}
D. Zhang, W. Lin, Z. Lin, L. Jia, W. Zheng, and F. Huang,  Lu$_2$O$_3$: A promising ultrawide bandgap semiconductor for deep UV photodetector, Appl. Phys. Lett. \textbf{118}, 211906 (2021).

\bibitem{OME-2020-AKIHIKO}
S. Matsumoto and A. Ito, Chemical vapor deposition route to transparent thick films of Eu$^{3+}$-doped HfO$_2$ and Lu$_2$O$_3$ for luminescent phosphors, Optical Materials Express, Opt. Mater. Express \textbf{10}, 899 (2020).

\bibitem{PR-1967-Lee}
R. S. Lee and S. Legvold, Hall Effect of Gadolinium, Lutetium, and Yttrium Single Crystals, Phys. Rev. \textbf{162}, 431 (1967).

\bibitem{JCP-1973-Sped}
F. H. Spedding and J. J. Croat, Magnetic properties of high purity yttrium, lanthanum, and lutetium and the effects of impurities on these properties, J. Chem. Phys. \textbf{59}, 2451 (1973).

\bibitem{Growth-2014-Guzik}
M. Guzik, J. Pejchal, A. Yoshikawa, A. Ito, T. Goto, M. Siczek, T. Lis, and G. Boulon, Structural Investigations of Lu$_2$O$_3$ as Single Crystal and Polycrystalline Transparent Ceramic, Cryst. Growth Des. \textbf{14}, 3327 (2014).

\bibitem{SIA-2014-AS}
A. G. Shard, Detection limits in XPS for more than 6000 binary systems using Al and Mg K$_\alpha$ X-rays, Surf. Interface Anal. \textbf{46}, 175 (2014).

\bibitem{JETP-2013-Kai}
V. V. Kaichev, T. I. Asanova, S. B. Erenburg, T. V. Perevalov, V. A. Shvets, and V. A. Gritsenko, Atomic and electronic structures of lutetium oxide Lu$_2$O$_3$,  J. Exp. Theor. Phys. \textbf{116}, 323 (2013).

\bibitem{Book-1993-AM}
J. A. Mydosh, Spin Glasses: An Experimental Introduction (Taylor \& Francis, London, 1993).
 
\bibitem{PRB-2009-Wang}
F. Wang, J. Kim, Y.-J. Kim, and G. D. Gu, Spin-glass behavior in LuFe$_2$O$_{4+\delta}$, Phys. Rev. B \textbf{80}, 024419 (2009).

\bibitem{JPCM-2008-Huang}
W.-G Huang, X.-Q Zhang, H.-F Du, R.-F Yang, Y.-K Tang, Y. Sun, and Z.-H Cheng, Intrinsic exchange bias effect in phase-separated La$_{0.82}$Sr$_{0.18}$CoO$_3$ single crystal, J. Phys.: Condens. Matter \textbf{20}, 445209 (2008).

\bibitem{Science-2020-Kam}
U. Kamber, A. Bergman, A. Eich, D. Iu\c{s}an, M. Steinbrecher,
N. Hauptmann, L. Nordstr\"{o}m, M. I. Katsnelson, D. Wegner, O. Eriksson, A. A. Khajetoorians, Self-induced spin glass state in elemental and crystalline neodymium, Science \textbf{368}, 966 (2020).

\bibitem{JPFMP-1975-PW}
S. F. Edwards and P. W. Anderson, Theory of spin glasses, J. Phys. F: Met. Phys. \textbf{5}, 965 (1975).

\bibitem{PRL-1984-Peter}
R. P. Peters, Ch. Buchal, M. Kubota, R. M. Mueller, and F. Pobell, Palladium-Iron: A Giant-Moment Spin-Glass at Ultralow Temperatures, Phys. Rev. Lett. \textbf{53}, 1108 (1984).

\bibitem{MCP-2010-Ben}
L. Ben Farhat, M. Amami, E.K. Hlil, and R. Ben Hassen, Synthesis, structure and magnetic properties of the Lu$_2$O$_3$–CoO mixed system, Mater. Chem. Phys. \textbf{123}, 737 (2010).



\end{thebibliography}
\end{document}